\documentclass[aps,twocolumn,showpacs,floatfix,superscriptaddress,letterpaper,longbibliography]{revtex4-1}
\usepackage{graphicx}
\usepackage{bm}
\usepackage{amsmath,amsfonts}
\usepackage{amsbsy}
\usepackage[utf8]{inputenc}
\DeclareUnicodeCharacter{03BC}{\mbox{$\mu$}}
\DeclareUnicodeCharacter{2009}{ }

\begin{document}

\title{Propagation and attenuation of sound in one-dimensional quantum
  liquids}

\author{K. A. Matveev}

\affiliation{Materials Science Division, Argonne National Laboratory,
  Argonne, Illinois 60439, USA}

\author{A. V. Andreev}

\affiliation{Department of Physics, University of Washington, Seattle,
  Washington 98195, USA}

\date{August 27, 2018}

\begin{abstract}

  At low temperatures, elementary excitations of a one-dimensional
  quantum liquid form a gas that can move as a whole with respect to
  the center of mass of the system.  This internal motion attenuates
  at exponentially long time scales.  As a result, in a broad range of
  frequencies the liquid is described by two-fluid hydrodynamics, and
  the system supports two sound modes.  The physical nature of the two
  sounds depends on whether the particles forming the quantum liquid
  have a spin degree of freedom.  For particles with spin, the modes
  are analogous to the first and second sound modes in superfluid
  $^4$He, which are the waves of density and entropy, respectively.
  When dissipative processes are taken into account, we find that at
  low frequencies the second sound is transformed into heat diffusion,
  while the first sound mode remains weakly damped and becomes the
  ordinary sound.  In a spinless liquid the entropy and density
  oscillations are strongly coupled, and the resulting sound modes are
  hybrids of the first and second sound.  As the frequency is lowered
  and dissipation processes become important, the crossover to
  single-fluid regime occurs in two steps.  First the hybrid modes
  transform into predominantly density and entropy waves, similar to
  the first and second sound, and then the density wave transforms to
  the ordinary sound and the entropy wave becomes a heat diffusion
  mode.  Finally, we account for the dissipation due to viscosity and
  intrinsic thermal conductivity of the gas of excitations, which
  controls attenuation of the sound modes at high frequencies.
\end{abstract}
\maketitle

\section{Introduction}
\label{sec:introduction}

At low frequencies vibrations of a fluid propagate through it in the
form of sound waves.  The latter are oscillations of density
accompanied by oscillations of entropy density, such that the entropy
per particle remains unchanged \cite{landau_fluid_2013}.  When liquid
$^4$He undergoes the superfluid transition, the physics of sound waves
changes dramatically.  Instead of a single sound wave, two kinds of
sound propagate in the superfluid.  The first sound is an oscillation
of mostly the particle density, whereas the second sound is a wave of
entropy \cite{landau_theory_1941}.  The physical origin of this
behavior is that the gas of elementary excitations can move with
respect to the rest of the superfluid without friction.  Thus
superfluid $^4$He can be treated as a system of two interpenetrating
fluids and described theoretically in the framework of two-fluid
hydrodynamics \cite{landau_theory_1941, khalatnikov_introduction_2000,
  donnelly_two-fluid_2009}.  The existence of two sound modes is a
generic property of superfluids; it was observed experimentally in
superfluid $^4$He \cite{donnelly_two-fluid_2009} and also more
recently in an ultracold Fermi gas with resonant interactions
\cite{sidorenkov_second_2013}.

In this paper we study sound in one-dimensional quantum liquids, such
as the electron liquid in long quantum wires
\cite{yacoby_nonuniversal_1996, scheller_possible_2014} or cold atomic
gases in narrow traps \cite{kinoshita_observation_2004,
  moritz_confinement_2005}.  These systems are commonly described
theoretically in the framework of the Tomonaga-Luttinger liquid theory
\cite{tomonaga_remarks_1950, haldane_luttinger_1981,
  haldane_effective_1981, giamarchi_quantum_2004}.  In the simplest
form of this theory the elementary excitations of a one-dimensional
quantum liquid are noninteracting bosons with linear spectrum.  To
account for the relaxation of the system to equilibrium, the formally
irrelevant perturbations must be added to the Luttinger liquid
Hamiltonian.

Sound waves propagate only in systems that relax to local thermal
equilibrium on time scales shorter than the period of oscillations of
density.  Relaxation of one-dimensional quantum liquids has been
studied in a number of recent papers
\cite{khodas_fermi-luttinger_2007, karzig_energy_2010,
  imambekov_one-dimensional_2012, ristivojevic_relaxation_2013,
  matveev_decay_2013, arzamasovs_kinetics_2014,
  protopopov_relaxation_2014}.  The rate of collisions $\tau_{\rm
  ex}^{-1}$ between the thermally excited quasiparticles scales as a
power law of the temperature $T$.  In particular, in spinless
Luttinger liquids $\tau_{\rm ex}^{-1}\propto T^7$
\cite{arzamasovs_kinetics_2014, protopopov_relaxation_2014}, whereas
for weakly interacting fermions with spin $\tau_{\rm ex}^{-1}\propto
T$ \cite{karzig_energy_2010}.

At time scales longer than $\tau_{\rm ex}$ the excitations of the
one-dimensional quantum liquid come to equilibrium with each other and
form a gas, which in general moves at some velocity $u_{\rm ex}$.  It
is important to keep in mind that $u_{\rm ex}$ does not necessarily
equal the velocity of the center of mass of the liquid.  The two
velocities do equilibrate, but this relaxation involves different
types of scattering processes, whose rate scales exponentially with
temperature, $\tau^{-1}\propto e^{-D/T}$
\cite{micklitz_transport_2010, matveev_scattering_2012,
  matveev_scattering_2014}, where $D$ is the bandwidth of the model.
Thus, at low temperatures there is a broad range of frequencies
\begin{equation}
  \label{eq:frequency_range} \tau^{-1} \ll \omega \ll \tau_{\rm
ex}^{-1},
\end{equation}
in which the gas of excitations moves with respect to the rest of the
fluid with negligible friction.  In this regime the behavior of a
one-dimensional quantum liquid is analogous to that of a superfluid.
In particular, the system is described by two-fluid hydrodynamics and
supports two sound modes \cite{matveev_second_2017,
  matveev_hybrid_2018}.  At $\omega \ll \tau^{-1}$ the two-fluid
behavior is destroyed, and only one sound mode remains.  This is a
manifestation of the fact that one-dimensional quantum liquid is not a
superfluid.

In this paper we expand on the earlier results
\cite{matveev_second_2017, matveev_hybrid_2018} by accounting for the
dissipation processes affecting the sound modes.  This enables us to
investigate the crossover from the two-fluid behavior in the frequency
range (\ref{eq:frequency_range}) to ordinary hydrodynamics at $\omega
\to0$.  The crossover at $\omega \sim \tau^{-1}$ is especially
interesting in the case of spinless quantum liquids, where the two
sound modes have an unusual hybrid nature \cite{matveev_hybrid_2018}.
In addition to including the slow relaxation at the time scale $\tau$,
we consider the effects of finite $\tau_{\rm ex}$, by including the
effects of viscosity and thermal conductivity in the hydrodynamic
equations.  This enables us to obtain the full picture of propagation
and attenuation of the sound modes in one-dimensional quantum liquids
at all frequencies $\omega \ll \tau_{\rm ex}^{-1}$.

The paper is organized as follows.  In Sec.~\ref{sec:thermodynamics}
we discuss the properties of bosonic elementary excitations of
interacting one-dimensional systems in the Luttinger liquid
approximation.  We obtain the basic thermodynamic characteristics of
the quantum liquid in the lowest order at $T\ll D$.  In
Sec.~\ref{sec:hydrodynamics} we derive the hydrodynamic equations of
the one-dimensional quantum liquid near equilibrium, while accounting
for both the fast and slow relaxation processes described by the rates
$\tau_{\rm ex}^{-1}$ and $\tau^{-1}$.  In Sec.~\ref{sec:sound_modes}
we solve the hydrodynamics equations in the frequency range
(\ref{eq:frequency_range}) and obtain the two sound modes discussed in
Refs.~\cite{matveev_second_2017, matveev_hybrid_2018}.  The crossover
from two sound modes to a single sound at $\omega\ll\tau^{-1}$ is
studied in Sec.~\ref{sec:crossover}.  The effect of the fast
relaxation processes on the attenuation of sound is studied in
Sec.~\ref{sec:attenuation_fast}.  We discuss our results in
Sec.~\ref{sec:discussion}.

\section{Thermodynamics of Luttinger liquids}
\label{sec:thermodynamics}

We start by reviewing the general properties of one-dimensional
quantum liquids within the Luttinger liquid theory.  Let us consider a
system of $N$ spinless particles confined to a region of length $L$
with periodic boundary conditions.  The Luttinger liquid theory
describes the low-energy properties of the system.  The elementary
excitations are bosons that occupy discrete states with momenta $k$
that are multiples of $2\pi\hbar/L$, excluding $k=0$.  The momentum of
the system is given by \cite{haldane_luttinger_1981, haldane_effective_1981}
\begin{equation}
  \label{eq:P}
  P=\frac{\pi\hbar}{L}NJ+\sum_k k N_{k}.
\end{equation}
Here $N_k$ is the occupation number of the quasiparticle state $k$,
while $J$ is an integer that is even if the particles of the quantum
liquid are bosons and has parity opposite to that of $N$ in the case
of fermions.  The first term in Eq.~(\ref{eq:P}) accounts for the
momentum of the moving liquid in the ground state defined by the
absence of excitations, $N_k=0$.  The second term in Eq.~(\ref{eq:P})
is the momentum of the bosonic excitations in the Luttinger liquid.

The number $J$ will play an important role in the hydrodynamic
description of the Luttinger liquid.  It describes the quantization of
momentum of the liquid in the ground state, which can be understood as
follows.  The smallest change of the momentum of the system that is
not accompanied by creation or destruction of quasiparticles is
obtained by adding momentum quantum $2\pi\hbar/L$ to each of the $N$
particles of the fluid.  This corresponds to $J\to J+2$ in
Eq.~(\ref{eq:P}).  For Galilean invariant systems considered in this
paper it is convenient to express the first term in Eq.~(\ref{eq:P})
as $mNu_0$, where $m$ is the mass of the particles, and the velocity
\begin{equation}
  \label{eq:u_0}
  u_0=\frac{\pi\hbar J}{mL}.
\end{equation}
One can think of $u_0$ as the velocity of the reference frame in which
$J=0$.  In the two-fluid hydrodynamics it will play the role of the
velocity of the superfluid component.

The energy of a Galilean invariant quantum liquid with the number of
particles close to a reference value $N_0$ in the Luttinger liquid
theory has the form \cite{haldane_luttinger_1981,
  haldane_effective_1981}
\begin{equation} 
 \label{eq:E}
  E=\frac{mv^2}{2N_0}(N-N_0)^2+\frac{\pi^2\hbar^2NJ^2}{2mL^2}
    +\sum_k\epsilon(k)N_{k}.
\end{equation}
Here $v$ is the velocity of the bosonic excitations, determined by the
compressibility of the quantum liquid in the ground state.  The energy
of the excitations is therefore usually written as $\epsilon(k)=v|k|$.
This expression applies to the case of a stationary liquid, i.e., at
$J=0$.  At nonvanishing $J$, the excitation energy
\begin{equation}
  \label{eq:epsilon}
  \epsilon(k)=v|k|+u_0k
\end{equation}
is found by the Galilean transformation to the frame moving with
velocity (\ref{eq:u_0}).

In addition to spinless quantum liquids, we will also consider a
one-dimensional system of spin-$\frac12$ fermions with repulsive
interactions.  This system can also be described by a Luttinger liquid
theory with a few modifications to account for the spin degrees of
freedom \cite{giamarchi_quantum_2004}.  The main difference with the
spinless case is that instead of one branch of bosonic excitations
there are two such branches.  The two types of bosons account for
excitations in the charge and spin channels and propagate at two
different speeds, $v_\rho$ and $v_\sigma$, respectively.  It is worth
mentioning that in the case of spin-$\frac12$ fermions in addition to
the parameters $N$ and $J$, the state of the Luttinger liquid 
depends on two similar quantum numbers in the spin channel.  In the absence
of magnetic field and other violations of spin rotation symmetry these
variables take zero values and do not affect sound in the system.

We now turn to the discussion of relaxation of one-dimensional quantum
liquids to thermal equilibrium.  Weak interactions between the
elementary excitations bring about their relaxation to an equilibrium
state.  These interactions belong to two classes.  The strongest
interactions are accounted for by the irrelevant perturbations to the
Hamiltonian of the Luttinger liquid \cite{giamarchi_quantum_2004}.
Their strength scales as a power law of temperature.  An important
feature of these interactions is that they conserve the total momentum
of the excitations given by the second term in Eq.~(\ref{eq:P}) and do
not change $J$.  As a result, at the time scale $\tau_{\rm ex}$ the
occupation numbers $N_k$ approach the equilibrium values
\begin{equation}
  \label{eq:Bose}
  N_{k}=\left[\exp\left(\frac{
        \epsilon(k)-u_{\rm ex}k}{T}\right)-1\right]^{-1}.
\end{equation}
The parameter $u_{\rm ex}$ is the velocity of the gas of excitations.
Conservation of momentum in the collisions of excitations means that
in the thermodynamic equilibrium $u_{\rm ex}$ does not have to vanish.

Interactions of the second type conserve the total momentum of the
system but not the two terms in the right hand side of
Eq.~(\ref{eq:P}) separately.  The resulting scattering processes are
analogous to the umklapp scattering of phonons in a crystal.  At low
temperatures the corresponding relaxation time is exponentially long,
$\tau\propto e^{D/T}$ \cite{micklitz_transport_2010,
  matveev_scattering_2012, matveev_scattering_2014}.  These processes
result in friction between the gas of excitations and the rest of the
quantum liquid, and can be expressed as relaxation of $u_{\rm
  ex}-u_0^{}$ to zero,
\begin{equation}
  \label{eq:tau_definition}
  \frac{d}{dt}(u_{\rm ex}-u_0^{}) = -\frac{u_{\rm ex}-u_0^{}}{\tau}.
\end{equation}
At long time scales $t\gg \tau$ both $u_{\rm ex}$ and $u_0^{}$
approach the center of mass velocity of the liquid.

In Sec.~\ref{sec:hydrodynamics} we apply the two-fluid hydrodynamic
theory to the description of one-dimensional quantum liquids.  This
approach is applicable at time scales that are long compared with
$\tau_{\rm ex}$, but not necessarily longer than $\tau$.  Under these
conditions the motion of the quantum liquid is characterized by two
velocities: $u_0$ and $u_{\rm ex}$.  To obtain the hydrodynamic
equations we will need expressions for a number of thermodynamic
properties of the Luttinger liquid at given density of particles $n$,
temperature, $u_0$, and $u_{\rm ex}$.  Since we will only use
linearized hydrodynamic equations, the expressions below are limited
to terms up to first order in $u_0$, $u_{\rm ex}$, and the deviation
of density $n$ from its reference value $n_0$.

Substituting the expression (\ref{eq:Bose}) for the occupation numbers
into Eq.~(\ref{eq:E}) we obtain the energy density
\begin{equation}
  \label{eq:varepsilon}
    \varepsilon=\frac{\pi T^2}{6\hbar {\widetilde v}}.
\end{equation}
We wrote Eq.~(\ref{eq:varepsilon}) in the form that applies to both
the spinless quantum liquid and to the case of spin-$\frac12$
fermions.  In the former case $\widetilde v=v$, whereas in the latter one
\begin{equation}
  \label{eq:tilde_v}
  {\widetilde v} = \left(\frac{1}{v_\rho}+\frac{1}{v_\sigma}\right)^{-1}.
\end{equation}
To obtain the entropy density $s$ we use the thermodynamic definition
of temperature $T=(\partial\varepsilon/\partial s)_n$ and find
\begin{equation}
  \label{eq:s}
  s=\frac{\pi T}{3\hbar\widetilde v}.
\end{equation}
We then rewrite the result (\ref{eq:varepsilon}) in an alternative
form
\begin{equation}
  \label{eq:varepsilon_vs_s}
    \varepsilon = \frac{3\hbar}{2\pi}{\widetilde v} s^2.
\end{equation}
Using the thermodynamic expression for the pressure $\Pi=-\varepsilon
+Ts +n(\partial\varepsilon/\partial n)_s$, we find
\begin{eqnarray}
  \label{eq:Pi}
  \Pi &=& \Pi^{(0)}
  + \varepsilon\frac{\partial_n {(n\widetilde v)}}{{\widetilde v}}.
\end{eqnarray}
Here $\Pi^{(0)}$ is the pressure at zero temperature.  It originates
from the ground state energy density omitted in
Eqs.~(\ref{eq:varepsilon}) and (\ref{eq:varepsilon_vs_s}) and cannot
be determined within the Luttinger liquid theory.

Similarly, substitution of Eqs.~(\ref{eq:epsilon}) and (\ref{eq:Bose})
into the expression (\ref{eq:P}) gives the momentum density
\begin{equation}
  \label{eq:p}
  p=mnu_0+\frac{2\varepsilon}{\bar v^2}(u_{\rm ex}-u_0).
\end{equation}
Here $\bar v=v$ in the case of spinless quantum liquid, while 
\begin{equation}
  \label{eq:v_bar}
  \bar v = \left(
           \frac{v_\rho^{-1}+v_\sigma^{-1}}{v_\rho^{-3}+v_\sigma^{-3}}
        \right)^{1/2}
\end{equation}
for spin-$\frac12$ fermions.

\section{Hydrodynamics of one-dimensional quantum liquids}
\label{sec:hydrodynamics}

As we saw in Sec.~\ref{sec:thermodynamics}, at frequencies $\omega \ll
\tau_{\rm ex}^{-1}$ the motion of a one-dimensional quantum liquid is
characterized by two velocities.  Velocity $u_0$ is associated with
the ground state motion of the liquid, whereas $u_{\rm ex}$ is the
velocity of the gas of excitations.  These two velocities are
analogous to the velocities of the superfluid and normal components of
superfluid $^4$He \cite{landau_theory_1941,
  khalatnikov_introduction_2000}.  In particular, the momentum density
(\ref{eq:p}) of the liquid can be written in the form
\begin{equation}
  \label{eq:momentum_density}
  p=\rho_0^{} u_0^{} +\rho_{\rm ex} u_{\rm ex}.
\end{equation}
Here the mass densities of the two components are given by
\begin{equation}
  \label{eq:rhos}
  \rho_0^{}=\rho -\rho_{\rm ex},
\quad
  \rho_{\rm ex}=\frac{\pi T^2}{3\hbar {\widetilde v}\bar v^2},
\end{equation}
with $\rho=mn$.

To fully describe the state of the one-dimensional fluid, in addition
to the velocities $u_0$ and $u_{\rm ex}$ one should specify the
particle density $n$ and temperature $T$.  Thus the motion of the
liquid is described by four evolution equations.  The first three of
these equations express the usual conservation laws for the mass,
energy, and momentum.  In this paper we limit ourselves to the effects
that are linear in small deviations from equilibrium.  In this
approximation energy dissipation is neglected, and energy conservation
is equivalent to conservation of entropy.  This yields continuity
equations for the mass, entropy, and momentum densities
\begin{subequations}
  \label{eq:continuity}
  \begin{eqnarray}
    \label{eq:continuity_n}
    \partial_t \rho + \partial_x j_\rho&=&0,
\\[1ex]
    \label{eq:continuity_entropy}
    \partial_t s + \partial_x j_s&=&0,
\\[1ex]
    \label{eq:continuity_p}
    \partial_t p + \partial_x j_p&=&0.
  \end{eqnarray}
Here $j$, $j_s$, and $j_p$ are the particle, entropy, and momentum
currents. 

Let us now show that the fourth evolution equation has the form
\begin{equation}
    \label{eq:continuity_u_0}
        \partial_t u_0 + \partial_x
        j_{u_0} = -\frac{\rho_{\rm ex}}{\rho}\,\frac{u_0-u_{\rm ex}}{\tau}.
\end{equation}
\end{subequations}
At $\tau\to\infty$ there is a fourth conserved quantity, namely the
quantum number $J$ of the given state of the Luttinger liquid.
Accordingly, Eq.~(\ref{eq:continuity_u_0}) takes the form of the
continuity equation for the density $J/L$, see Eq.~(\ref{eq:u_0}).  We
note that at $\tau\to\infty$ this equation is equivalent to the
equation describing the time evolution of superfluid velocity in
two-fluid hydrodynamic theory of liquid $^4$He
\cite{landau_theory_1941, khalatnikov_introduction_2000}.

At finite $\tau$ the quantum number $J$ is no longer conserved,
resulting in the friction between the two components of the fluid, see
Eq.~(\ref{eq:tau_definition}).  In this case the rate of change of
velocity $u_0$ cannot be fully expressed as a gradient of the current
$j_{u_0}$ in Eq.~(\ref{eq:continuity_u_0}).  Since the dissipative
processes that result in friction conserve the momentum
(\ref{eq:momentum_density}), the rate $du_0/dt$ in the right-hand side
of Eq.~(\ref{eq:continuity_u_0}) can be found by combining
Eq.~(\ref{eq:tau_definition}) with $\rho_0 (du_0/dt) +\rho_{\rm ex}
(du_{\rm ex}/dt)=0$.

Because of the Galilean invariance, the mass current $j_\rho$ in
Eq.~(\ref{eq:continuity_n}) equals the momentum density,
\begin{subequations}
\label{eq:currents}
\begin{equation}
  \label{eq:j}
  j_\rho=\rho_0^{} u_0 +\rho_{\rm ex} u_{\rm ex}.
\end{equation}
The expressions for the remaining three currents are well understood
in the theory of superfluidity of liquid $^4$He
\cite{khalatnikov_introduction_2000}.  Adapting them to one dimension,
we obtain
\begin{eqnarray}
  \label{eq:j_s}
  j_s&=&su_{\rm ex}-\frac{\kappa_{\rm ex}}{T}\partial_x T,
\\
  \label{eq:j_p}
  j_p&=&\Pi-\zeta_1\partial_x(p-\rho\:\! u_{\rm ex})-\zeta_2\partial_xu_{\rm
    ex},
\\
  \label{eq:j_u_0}
  j_{u_0}&=&\frac{\mu}{m}-\zeta_3\partial_x(p-\rho\:\! u_{\rm ex})
                     -\zeta_4\partial_xu_{\rm ex},
\end{eqnarray}
\end{subequations}
where $\mu$ is the chemical potential of the quantum liquid.
Expressions (\ref{eq:j_s})--(\ref{eq:j_u_0}) include the dissipative
contributions to the respective currents, which are parametrized by
four coefficients of bulk viscosity $\zeta_1,\ldots,\zeta_4$ and the
thermal conductivity of the gas of excitations $\kappa_{\rm ex}$.  We
note that these parameters are proportional to $\tau_{\rm ex}$ and
that the viscosity coefficients satisfy an Onsager relation
$\zeta_1=\zeta_4$.

\section{Two sound modes}
\label{sec:sound_modes}

Let us now consider the solutions of the hydrodynamic equations
(\ref{eq:continuity}) and (\ref{eq:currents}) for frequencies in the
range (\ref{eq:frequency_range}), i.e., assuming $\tau\to\infty$ and
$\tau_{\rm ex}=0$.  The former limit enables us to set the right-hand
side of Eq.~(\ref{eq:continuity_u_0}) to zero.  In addition, since the
bulk viscosities and the thermal conductivity of the gas of
excitations are proportional to $\tau_{\rm ex}$, we will set
$\zeta_i=0$ for all $i$ and $\kappa_{\rm ex}=0$ in
Eqs.~(\ref{eq:j_s})--(\ref{eq:j_u_0}).  Following
Ref.~\cite{khalatnikov_introduction_2000}, instead of the entropy
density $s$ we will use the entropy per unit mass $\sigma=s/\rho$.
The resulting four equations are
\begin{subequations}
  \label{eq:dissipationless}
  \begin{eqnarray}
    \label{eq:dissipationless_rho}
    \partial_t \rho + \rho_0^{}\, \partial_x u_0 
     + \rho_{\rm ex}\, \partial_x u_{\rm ex} &=& 0,
\\
    \label{eq:dissipationless_sigma}
    \partial_t \sigma 
    + \frac{\rho_0^{}}{\rho}\sigma\,\partial_x (u_{\rm ex}-u_0) &=& 0,
\\
    \label{eq:dissipationless_p}
    \rho_0^{}\,\partial_t u_0 + \rho_{\rm ex}\, \partial_t u_{\rm ex} 
     + \partial_x \Pi &=& 0,
\\
    \label{eq:dissipationless_u_0}
    \partial_t u_0 + \frac{1}{m}\partial_x \mu &=& 0.
\end{eqnarray}
\end{subequations}
The four first order differential equations (\ref{eq:dissipationless})
can be easily reduced to two second order equations for $\rho$ and
$\sigma$.  First, combining Eqs.~(\ref{eq:dissipationless_rho}) and
(\ref{eq:dissipationless_p}), we obtain
\begin{equation}
  \label{eq:rho_eqn}
  \partial_t^2\rho = \partial_x^2 \Pi.
\end{equation}
Then, using the thermodynamic relation $d\;\!\Pi-n\;\!d\mu=s\;\!dT$ and
excluding $u_0$ and $u_{\rm ex}$ from
Eqs.~(\ref{eq:dissipationless_sigma})--(\ref{eq:dissipationless_u_0}),
we find
\begin{equation}
  \label{eq:sigma_eqn}
  \partial_t^2\sigma = \frac{\rho_0^{}}{\rho_{\rm ex}}\sigma^2\partial_x^2 T.
\end{equation}
In Eqs.~(\ref{eq:rho_eqn}) and (\ref{eq:sigma_eqn}) we will choose
$\rho$ and $\sigma$ as two variables describing the state of the
fluid, and treat the pressure and temperature as their functions
$\Pi(\rho,\sigma)$ and $T(\rho,\sigma)$.

To study the sound modes in the system, we assume that $\rho$ and
$\sigma$ have small variations of the form $\delta\rho\cos[q(x-ct)]$
and $\delta\sigma\cos[q(x-ct)]$.  Expanding Eq.~(\ref{eq:rho_eqn}) and
(\ref{eq:sigma_eqn}) to linear order in $\delta\rho$ and
$\delta\sigma$, we get
\begin{subequations}
  \label{eq:sound_eqns_linearized}
\begin{eqnarray}
  \label{eq:rho_eqn_linearized}
  \big(c^2-A_{11}\big)\delta\rho-A_{12}\,\delta\sigma&=&0,
\\[1ex]
  \label{eq:sigma_eqn_linearized}
  -A_{21}\,\delta\rho+\big(c^2-A_{22}\big)\delta\sigma&=&0.
\end{eqnarray}
\end{subequations}
Here the coefficients $A_{ij}$ are given by
\begin{subequations}
  \label{eq:A}
\begin{eqnarray}
  \label{eq:A11}
  A_{11}&=&\left(\frac{\partial\Pi}{\partial\rho}\right)_\sigma,
\\
  \label{eq:A12}
  A_{12}&=&\left(\frac{\partial\Pi}{\partial\sigma}\right)_\rho,
\\
  \label{eq:A21}
  A_{21}&=&\frac{\rho_0^{}}{\rho_{\rm ex}}\sigma^2
          \left(\frac{\partial T}{\partial\rho}\right)_\sigma,
\\
  \label{eq:A22}
  A_{22}&=&\frac{\rho_0^{}}{\rho_{\rm ex}}\sigma^2
          \left(\frac{\partial T}{\partial\sigma}\right)_\rho.
\end{eqnarray}
\end{subequations}
The system of equations (\ref{eq:sound_eqns_linearized}) has
nonvanishing solutions with velocities given by
\begin{equation}
  \label{eq:v^2}
  c_\pm^2=\frac{A_{11}+A_{22}}{2}
    \pm\frac12\sqrt{(A_{11}-A_{22})^{2}+4A_{12}A_{21}}.
\end{equation}
Thus, in the frequency range (\ref{eq:frequency_range}) the one
dimensional quantum liquid supports two sound modes.

For a system of spin-$\frac12$ fermions, using
Eqs.~(\ref{eq:s})--(\ref{eq:Pi}) we obtain the following results for
$A_{ij}$
\begin{subequations}
  \label{eq:A_result}
\begin{eqnarray}
  \label{eq:A11_result}
  A_{11}&=& v_\rho^2 + \frac{\pi T^2}{6\hbar\rho\widetilde v^2}
                     \partial_\rho^2\big(\rho^2\widetilde v\big),
\\
  \label{eq:A12_result}
  A_{12}&=& \frac{\rho T}{\widetilde v}\partial_\rho(\rho \widetilde v),
\\
  \label{eq:A21_result}
  A_{21}&=&\frac{\pi\bar v^2T}{3\hbar\rho^2\widetilde v^2}
          \partial_\rho(\rho \widetilde v),
\\
  \label{eq:A22_result}
  A_{22}&=&\bar v^2.
\end{eqnarray}
\end{subequations}
In Eq.~(\ref{eq:A11_result}) we used the definition of velocity
$v_\rho$ in terms of the zero temperature compressibility,
$v_\rho^2=\partial_\rho\Pi^{(0)}$.  Note that within the simple
Luttinger liquid theory based on Eqs.~(\ref{eq:P}) and (\ref{eq:E})
only $A_{11}$ can be obtained with accuracy beyond the leading order
at $T\to0$.

Assuming the interactions between the fermions are repulsive, we have
$v_\rho> v_\sigma$, and thus $v_\rho>\bar v$, see
Eq.~(\ref{eq:v_bar}).  When the temperature approaches zero
Eq.~(\ref{eq:A_result}) yields $A_{11}=v_\rho^2$, $A_{22}=\bar v^2$,
$A_{12}=A_{21}=0$, and the two sound velocities (\ref{eq:v^2}) are
\begin{equation}
  \label{eq:velocities_w_spins}
  c_+=v_\rho,
\quad
  c_-=\bar v.
\end{equation}
In this limit, the faster and slower sound modes are pure waves of
density $\rho$ and entropy $\sigma$, respectively.  They are
completely analogous to the first and second sound in superfluid
$^4$He \cite{landau_theory_1941, khalatnikov_introduction_2000}.  At
small but finite temperature the nature of the sound modes remains
largely the same, but there is a small mixing of the oscillations of
$\rho$ and $\sigma$.  For example, the density oscillation
$\delta\rho$ of the first sound mode is accompanied by a weak
oscillation of entropy, $\delta\sigma\simeq [A_{21}/(v_\rho^2-\bar
v^2)]\delta\rho$.  This is in contrast with the usual adiabatic sound
in classical fluids, for which $\delta\sigma=0$ at any temperature.

In a spinless Luttinger liquid the nature of the two sound modes is
qualitatively different.  To obtain $A_{ij}$ in this case, one should
substitute $v_\rho=\widetilde v=\bar v=v$ into
Eq.~(\ref{eq:A_result}).  At $T\to0$ this yields $A_{11}=A_{22}=v^2$
and $A_{12}=A_{21}=0$.  As a result, the two sound velocities
(\ref{eq:v^2}) are equal, $c_\pm=v$.  At a small but finite
temperature, the leading order correction to $A_{22}$ is quadratic in
$T$ \cite{matveev_hybrid_2018}, as is the correction to $A_{11}$ in
Eq.~(\ref{eq:A11_result}).  Thus the term $4A_{12}A_{21}\propto T^2$
dominates the square root in Eq.~(\ref{eq:v^2}), resulting in a linear
in temperature splitting of the sound velocities:
\begin{equation}
  \label{eq:v_pm}
  c_\pm=v\pm\frac{\sqrt{A_{12}A_{21}}}{2v}
       =v\pm\sqrt{\frac{\pi T^2}{12\hbar\rho v^3}}\,\partial_\rho(\rho v).
\end{equation}
The sound mode propagating at the higher speed $c_+$ corresponds to
in-phase oscillations of density and entropy.  The relative magnitude
of the oscillations $\delta\rho/\delta\sigma=\sqrt{A_{12}/A_{21}}$
approaches a finite limit at $T\to0$.  The mode propagating at the
slower speed $c_-$ is characterized by oscillations of $\rho$ and
$\sigma$ that are of opposite signs.  The two sound modes are
therefore different from the usual first and second sounds, but
instead combine the features of both.  

This hybrid nature of the sound modes in a spinless quantum liquid was
discussed in Ref.~\cite{matveev_hybrid_2018}.  We note that in a
three-dimensional weakly interacting Bose gas the speeds of the first
and second sounds become equal at a certain temperature $T_1$
\cite{lee_low-temperature_1959}.  This results in hybridization of
these sound modes at temperatures in a narrow vicinity of $T_1$
\cite{verney_hybridization_2015}.  In contrast, in a one-dimensional
spinless quantum liquid strong hybridization of sound modes occurs at
any interaction strength, as long as $T\ll D$.

\section{Crossover between the regimes of one and two sound modes}
\label{sec:crossover}

The two sound modes discussed in Sec.~\ref{sec:sound_modes} exist in
the frequency range (\ref{eq:frequency_range}).  Under these
conditions the one-dimensional fluid behaves similarly to a superfluid
and can be thought of as consisting of two components moving with two
different velocities $u_0$ and $u_{\rm ex}$.  In contrast to
superfluids, there is friction between the two components of the
fluid, but it is negligible at $\omega\gg\tau^{-1}$.  On the other
hand, at low frequencies $\omega\ll\tau^{-1}$ the friction between the
two components is important, and one should expect equilibration of
the two velocities, $u_0=u_{\rm ex}$.  In this limit the
one-dimensional quantum liquid loses its superfluid properties and
behaves as an ordinary fluid.  In particular, only a single sound mode
is present at $\omega\ll\tau^{-1}$.  In this section we discuss the
crossover between the regimes with one and two sound modes at
$\omega\sim\tau^{-1}$.

The hydrodynamic equations (\ref{eq:continuity}) and
(\ref{eq:currents}) are applicable in this regime, with the only
limitation on the frequency being $\omega\ll\tau_{\rm ex}^{-1}$.
In contrast to our discussion in Sec.~\ref{sec:sound_modes}, one can no longer
assume $\tau\to\infty$.  On the other hand, one can still consider the
limit $\tau_{\rm ex}\to 0$ and neglect the bulk viscosities $\zeta_i$
and $\kappa_{\rm ex}$ in Eq.~(\ref{eq:currents}).  Under these
circumstances the hydrodynamic equations (\ref{eq:continuity}) can
again be rewritten in the form (\ref{eq:dissipationless}), but with
Eq.~(\ref{eq:dissipationless_u_0}) replaced by
\begin{equation}
  \label{eq:u_0_with_tau}
    \partial_t u_0 + \frac{1}{m}\partial_x \mu 
      = -\frac{\rho_{\rm ex}}{\rho}\,\frac{u_0-u_{\rm ex}}{\tau}.
\end{equation}
This change does not affect the derivation of Eq.~(\ref{eq:rho_eqn}),
but Eq.~(\ref{eq:sigma_eqn}) is replaced by
\begin{equation}
  \label{eq:sigma_eqn_with_tau}
  \partial_t^2\sigma +\frac{1}{\tau}\partial_t\sigma
      = \frac{\rho_0^{}}{\rho_{\rm ex}}\sigma^2\partial_x^2 T.
\end{equation}
Assuming that the small variations of $\rho$ and $\sigma$ in
Eqs.~(\ref{eq:rho_eqn}) and (\ref{eq:sigma_eqn_with_tau}) have the
forms $\delta\rho\, e^{i(qx-\omega t)}$ and $\delta\sigma\,
e^{i(qx-\omega t)}$, respectively, we find
\begin{subequations}
  \label{eq:sound_eqns_linearized_with_tau}
\begin{eqnarray}
  \label{eq:rho_eqn_linearized_with_tau}
  \big(\omega^2-A_{11}q^2\big)\,\delta\rho-A_{12}q^2\,\delta\sigma&=&0,
\\[1ex]
  \label{eq:sigma_eqn_linearized_with_tau}
  -A_{21}q^2\,\delta\rho
    +\left(\omega^2+\frac{i\omega}{\tau}-A_{22}q^2\right)\delta\sigma&=&0.
\end{eqnarray}
\end{subequations}
This system of linear equations has nontrivial solutions if
\begin{equation}
  \label{eq:determinant_with_tau}
  \big(\omega^2-A_{11}q^2\big)\left(\omega^2+\frac{i\omega}{\tau}-A_{22}q^2\right)
  -A_{12}A_{21}q^4=0.
\end{equation}
Equation (\ref{eq:determinant_with_tau}) enables one to study
propagation of perturbations of both density and entropy for arbitrary
$\omega\tau$, provided $\omega\tau_{\rm ex}\to0$.

\subsection{Low frequency regime}
\label{sec:low_frequency_regime}

We first consider the low-frequency limit $\omega\tau\to0$.  In this
case Eq.~(\ref{eq:determinant_with_tau}) takes the form
\begin{equation}
  \label{eq:determinant_with_large_tau}
  (\omega^2-A_{11}q^2\big)\left(\frac{i\omega}{\tau}-A_{22}q^2\right)
  -A_{12}A_{21}q^4=0.
\end{equation}
This equation for $\omega(q)$ has two types of solutions.  The first
one corresponds to the ordinary sound in the fluid.  For the
right-moving wave we get
\begin{equation}
  \label{eq:sound_solution}
  \omega=cq-i\tau\frac{A_{12}A_{21}}{2A_{11}}q^2.
\end{equation}
Here
\begin{equation}
  \label{eq:sound_velocity}
  c=\sqrt{A_{11}},
\end{equation}
which in combination with Eq.~(\ref{eq:A11}) gives the usual
expression for the speed of sound in terms of the adiabatic
compressibility \cite{landau_fluid_2013}.  The presence of the
imaginary part of $\omega$ in Eq.~(\ref{eq:sound_solution}) means that
the sound wave $\delta\rho\, e^{i(qx-\omega t)}$ gradually decays over
time.

The second solution of Eq.~(\ref{eq:determinant_with_large_tau}) is
purely imaginary,
\begin{equation}
  \label{eq:diffusive_solution}
  \omega=-i\tau \frac{A_{11}A_{22}-A_{12}A_{21}}{A_{11}}q^2.
\end{equation}
It shows that the second sound cannot exist at $\omega\ll\tau^{-1}$.
Instead, Eq.~(\ref{eq:diffusive_solution}) describes diffusive
propagation of heat in the system.  

In a system with thermal conductivity $\kappa$ and specific heat at
constant pressure $c_p$ the dependence $\omega(q)$ has the form
$\omega=-i(\kappa/c_p)q^2$.  This enables one to obtain the expression
for the thermal conductivity of the quantum liquid
\begin{equation}
  \label{eq:kappa_general}
  \kappa=\tau  A_{22}c_v.
\end{equation}
Here we used the thermodynamic relation
\begin{equation}
  \label{eq:c_p/c_v}
  \frac{c_p}{c_v}=\frac{A_{11}A_{22}}{A_{11}A_{22}-A_{12}A_{21}}
\end{equation}
with $A_{ij}$ given by Eq.~(\ref{eq:A}) to express $\kappa$ in terms
of the specific heat at constant density $c_v$.  In a Luttinger liquid
at low temperature, the entropy density (\ref{eq:s}) is a linear
function of $T$, resulting in $c_v=T(\partial s/\partial T)_\rho=s$.
We therefore conclude from Eqs.~(\ref{eq:kappa_general}) and
(\ref{eq:A22_result}) that the thermal conductivity of a
one-dimensional quantum liquid is given by
\begin{equation}
  \label{eq:kappa}
  \kappa=\frac{\pi T \bar v^2 \tau}{3\hbar\tilde v},
\end{equation}
where we used the result (\ref{eq:A22_result}) for $A_{22}$ at
$T\to0$.  The known result for the thermal conductivity of a spinless
one-dimensional quantum liquid \cite{degottardi_electrical_2015} is
obtained from Eq.~(\ref{eq:kappa}) by substitution $\bar v=\widetilde
v=v$.  It is worth mentioning that with the aid of
Eqs.~(\ref{eq:kappa_general}) and (\ref{eq:c_p/c_v}) the expression
(\ref{eq:sound_solution}) for the frequency of sound can be brought to
the form
\begin{equation}
  \label{eq:sound_solution_standard}
  \omega=cq-\frac{i\kappa}{2}\left(\frac{1}{c_v}-\frac{1}{c_p}\right)q^2.
\end{equation}
The last term in Eq.~(\ref{eq:sound_solution_standard}) reproduces the
well-known expression for the rate of sound absorption caused by the
finite thermal conductivity of the fluid \cite{landau_fluid_2013}.

\subsection{High frequency regime}
\label{sec:high_frequency_regime}

In the limit of large frequencies, $\omega\tau\to\infty$,
Eq.~(\ref{eq:determinant_with_tau}) has solutions of the form $\omega
=c_\pm q$ \cite{onefootnote} with velocities given by
Eq.~(\ref{eq:v^2}), and thus reproduces the results of
Sec.~\ref{sec:sound_modes}.  We now obtain the attenuation rate of
these sound modes in the first order in $(\omega\tau)^{-1}$.  A
straightforward calculation yields the solutions of
Eq.~(\ref{eq:determinant_with_tau}) in the form
\begin{equation}
  \label{eq:attenuation_high_omega}
  \omega=c_\pm q -\frac{i}{4\tau}
         \left[
           1\mp\frac{|A_{11}-A_{22}|}{\sqrt{(A_{11}-A_{22})^{2}+4A_{12}A_{21}}}
         \right].
\end{equation}
This expression gives very different attenuation rates of the sound
modes in the cases of liquids of spinless particles and spin-$\frac12$
fermions.

As discussed in Sec.~\ref{sec:sound_modes}, in a system with spins we
have $A_{12}A_{21}\ll(A_{11}-A_{22})^{2}$.  In this regime we obtain
the damping of the first and second sound modes in the form
\begin{subequations}
  \label{eq:attenuation_high_omega_spins}
  \begin{eqnarray}
    \label{eq:attenuation_high_omega_first_sound}
    \omega-v_\rho \;\!q&=& -\frac{i}{2\tau}
                    \frac{A_{12}A_{21}}{(A_{11}-A_{22})^{2}}
    \nonumber\\
             &=& -\frac{i}{\tau}\frac{\pi\bar v^2
               [\partial_\rho(\rho\widetilde v)]^2T^2}
              {6\hbar\rho\widetilde v^3(v_\rho^2-\bar v^2)^2},
    \\
    \label{eq:attenuation_high_omega_second_sound}
    \omega-\bar v\;\! q&=& -\frac{i}{2\tau}.
  \end{eqnarray}
\end{subequations}
The attenuation rate of the first sound is smaller than that of the
second sound by a parameter of order $(T/D)^2$.  In the case of a
spinless quantum liquid we have $A_{12}A_{21}\gg(A_{11}-A_{22})^{2}$,
and Eq.~(\ref{eq:attenuation_high_omega}) yields
\begin{equation}
  \label{eq:attenuation_high_omega_hybrid}
  \omega-c_\pm q = -\frac{i}{4\tau}.
\end{equation}
Thus the two hybrid sound modes decay at the same rate.

The qualitatively different behavior of the attenuation rates of the
different sound modes can be interpreted as follows.  Thermal
conductivity gives rise to dissipation in systems with non-uniform
temperature.  As discussed in Sec.~\ref{sec:sound_modes}, the second
sound mode in a system with spins is an almost pure wave of entropy
and, therefore, temperature.  Thus the second sound is attenuated
quite effectively
[Eq.~(\ref{eq:attenuation_high_omega_second_sound})].  In contrast,
the first sound is primarily a wave of density, with only a weak
disturbance in temperature.  As a result the processes of thermal
conductivity attenuate the first sound less effectively
[Eq.~(\ref{eq:attenuation_high_omega_first_sound})].  In a spinless
system, the two hybrid sounds carry equal in magnitude oscillations of
entropy, resulting in equal dissipation, see
Eq.~(\ref{eq:attenuation_high_omega_hybrid}).

\subsection{Crossover regime in a quantum liquid of spin-$\frac12$ fermions}
\label{sec:crossover_spins}

We begin the consideration of the crossover regime $\omega\tau\sim1$
with the first sound.  From our results (\ref{eq:sound_solution}) and
(\ref{eq:attenuation_high_omega_first_sound}) in the limiting cases or
small and large $\omega\tau$, we expect the attenuation to remain
small as $(T/D)^2$ at $\omega\tau\sim1$.  We therefore substitute
$\omega=cq+\delta\omega$ into Eq.~(\ref{eq:determinant_with_tau})
and linearize it in small $\delta \omega$.  This yields
\begin{equation}
  \label{eq:delta_omega}
  \delta\omega=\frac{A_{12}A_{21}}{2\sqrt{A_{11}}}
               \frac{q^2}{(A_{11}-A_{22})q+\frac{i}{\tau}\sqrt{A_{11}}}.
\end{equation}
Taking into account Eq.~(\ref{eq:A_result}), it is easy to see that
for $v_\rho>\bar v$ the correction (\ref{eq:delta_omega}) is indeed
small, $|\delta\omega|/\omega\sim (T/D)^2$.  The rate of decay of the
first sound is given by the imaginary part of $\omega$,
\begin{equation}
  \label{eq:Im_omega_first_sound}
  {\rm Im\,}\omega=-\frac{q^2\tau}{2}
              \frac{A_{12}A_{21}}{(q\tau)^2(A_{11}-A_{22})^2+A_{11}}.
\end{equation}
In the limits $q\to0$ and $q\to\infty$
Eq.~(\ref{eq:Im_omega_first_sound}) reproduces our earlier results
(\ref{eq:sound_solution}) and
(\ref{eq:attenuation_high_omega_first_sound}), respectively.  We note
also that ${\rm Re\,}\delta\omega\propto q$ at large $q$.  Thus the
speed of the first sound $c_+$ is slightly different from the speed
$c=\sqrt{A_{11}}$ of the ordinary sound,
\begin{equation}
  \label{eq:delta_c_first_sound}
  c_+=c+\frac{A_{12}A_{21}}{2\sqrt{A_{11}}(A_{11}-A_{22})}.
\end{equation}
Using Eq.~(\ref{eq:A_result}) it is easy to see that at low
temperature $|c-v_\rho|\sim |c_+-v_\rho|\sim v_\rho(T/D)^2$.  More
generally, the result (\ref{eq:delta_c_first_sound}) for the speed of
the first sound holds at frequencies in the range
(\ref{eq:frequency_range}) and can be obtained directly from
Eq.~(\ref{eq:v^2}).

To study the crossover from the second sound at $\omega\tau\gg 1$ to
diffusive heat transport at $\omega\tau\ll 1$ we rewrite
Eq.~(\ref{eq:determinant_with_tau}) in the form
\begin{equation}
\label{eq:determinant_alternative}
  \omega^2+\frac{i\omega}{\tau}-A_{22}q^2\left(1
  +\frac{A_{12}A_{21}q^2}{A_{22}\big(\omega^2-c^2q^2\big)}\right)=0.
\end{equation}
At the crossover we have $|\omega-cq|\sim \tau^{-1}$.  Then, using
Eq.~(\ref{eq:A_result}) we estimate the second term in parentheses of
Eq.~(\ref{eq:determinant_alternative}) to be of the order of
$(T/D)^2\ll1$.  Neglecting it and solving the resulting quadratic
equation, we obtain
\begin{equation}
  \label{eq:crossover_second_sound}
  \omega=\pm\sqrt{(\bar v q)^2-\frac{1}{(2\tau)^2}}-\frac{i}{2\tau},
\end{equation}
where we also used Eq.~(\ref{eq:A22_result}).  At $\bar
vq\gg\tau^{-1}$ Eq.~(\ref{eq:crossover_second_sound}) describes the
second sound with attenuation ${\rm Im\,}\omega=(2\tau)^{-1}$, in
agreement with Eq.~(\ref{eq:attenuation_high_omega_second_sound}).
This solution exists at all $\bar vq>(2\tau)^{-1}$, with real part
vanishing at $\bar vq=(2\tau)^{-1}$.  As smaller $q$ the frequencies
are imaginary.  At $q\to 0$ one of these damped modes becomes
$\omega=-i\tau(\bar v q)^2$, thereby transforming into the heat
diffusion mode (\ref{eq:diffusive_solution}).

The evolution of the first and second sound through the crossover at
$\omega\tau\sim1$ can be summarized as follows.  The first sound mode
does not experience significant changes at the crossover and gradually
evolves into the ordinary sound at $\omega\tau\ll1$.  Its attenuation
is always small, ${\rm Im\,}\omega\ll {\rm Re\,}\omega$.  In contrast,
the second sound mode experiences a sharp transition when its
wavevector $q$ crosses $\bar q=1/2\bar v\tau$.  The wave-like
propagation of heat, associated with the second sound, occurs at
$q>\bar q$, with damping becoming strong at $q\sim \bar q$.  At
$q<\bar q$ wave-like behavior is absent and at $q\ll \bar q$ the heat
propagation becomes diffusive.

\subsection{Crossover regime in a spinless quantum liquid}
\label{sec:crossover_spinless}

The crossover from two hybrid sound modes in a spinless quantum liquid
to a single sound at $\omega\tau\to0$ is more complicated.  We start
by noticing that as discussed in Sec.~\ref{sec:sound_modes}, at low
temperatures the difference between $A_{11}$ and $A_{22}$ in a
spinless system is negligible, and one can set $A_{11}=A_{22}=v$.
Then Eq.~(\ref{eq:determinant_with_tau}) can be rewritten in the form
\begin{equation}
  \label{eq:determinant_with_tau_spinless}
  \big(\omega^2-v^2q^2\big)^2 +\frac{i\omega}{\tau}\left(\omega^2-v^2q^2\right)
  -(v\:\!\delta c)^2q^4=0,
\end{equation}
Here we have introduced the difference of velocities of the hybrid
sound modes,
\begin{equation}
  \label{eq:delta_c}
  \delta c=c_+-c_-=\frac{\sqrt{A_{12}A_{21}}}{v},
\end{equation}
cf.~Eq~(\ref{eq:v_pm}).  We treat
Eq.~(\ref{eq:determinant_with_tau_spinless}) as a quadratic equation
for $\omega^2-v^2q^2$ and write the two solutions in the form
\begin{equation}
  \label{eq:quadratic_eq_solution}
  \omega^2-v^2q^2=-\frac{i\omega}{2\tau}
                  \pm\sqrt{-\left(\frac{\omega}{2\tau}\right)^2
                             +(v\:\!\delta c)^2q^4}.
\end{equation}
This expression is equivalent to
Eq.~(\ref{eq:determinant_with_tau_spinless}), and can be further
analyzed by an appropriate approximation of $\omega$ in the right-hand
side.

At $vq\gg\tau^{-1}$ one can approximate $\omega=vq$ in the right-hand
side of Eq.~(\ref{eq:quadratic_eq_solution}) and obtain
\begin{equation}
  \label{eq:hybrid_vs_dissipation}
  \omega-vq=-\frac{i}{4\tau}
             \pm\frac12\sqrt{-\frac{1}{(2\tau)^2}
                             +(\delta c)^2q^2}.
\end{equation}
In the limit $q\to\infty$ Eq.~(\ref{eq:hybrid_vs_dissipation})
recovers our earlier result (\ref{eq:attenuation_high_omega_hybrid}).
This asymptotic behavior requires $q\gg q^*$, where
\begin{equation}
  \label{eq:q^*}
  q^*=\frac{1}{2\tau\delta c}.
\end{equation}
Note that because of the smallness of the difference of velocities
$\delta c\ll v$ in the spinless case, at $q\sim q^*$ we have $\omega
\tau\gg 1$.  The attenuation rate remains constant, ${\rm
  Im\,}\omega=(4\tau)^{-1}$, for all $q>q^*$ \cite{twofootnote}, while
${\rm Re\,}\omega$ has a small nonlinear correction.  At $q<q^*$ the
real part ${\rm Re\,}\omega=vq$, whereas the attenuation of the two
sound modes is different,
\begin{equation}
  \label{eq:hybrid_vs_dissipation_below_q*}
  \omega-vq=\frac{i}{4\tau}
             \left[-1
             \pm\sqrt{1-(2\tau\delta cq)^2}
             \right].
\end{equation}

At $q\ll q^*$ the mode corresponding to the plus sign in
Eq.~(\ref{eq:hybrid_vs_dissipation_below_q*}) becomes weakly damped,
\begin{equation}
  \label{eq:spinless_first_sound_damping}
  \omega-vq=-\frac{i}{2}\tau(\delta cq)^2.
\end{equation}
For this mode $|\omega-vq|\ll |vq|$ for all $q$.  Thus the
approximation $\omega=vq$ leading to
Eq.~(\ref{eq:hybrid_vs_dissipation}) remains applicable even at
$vq\lesssim\tau^{-1}$.  Indeed, since the speed of sound $c=v$ in the
spinless quantum liquid, at $q\to0$
Eq.~(\ref{eq:spinless_first_sound_damping}) recovers the attenuation
(\ref{eq:sound_solution}) of the ordinary sound.  Thus this mode
behaves similarly to the first sound mode as discussed in
Sec.~\ref{sec:crossover_spins}. 

The sound mode corresponding to the minus sign in
Eq.~(\ref{eq:hybrid_vs_dissipation_below_q*}) behaves very
differently.  Its attenuation ${\rm Im\,}\omega\sim\tau^{-1}$ is small
compared to $vq$ only at $vq\gg\tau^{-1}$.  Thus at
$vq\lesssim\tau^{-1}$ Eqs.~(\ref{eq:hybrid_vs_dissipation}) and
(\ref{eq:hybrid_vs_dissipation_below_q*}) are inapplicable.  Instead,
in this regime one can neglect the last term in
Eq.~(\ref{eq:determinant_with_tau_spinless}), resulting in
$\omega^2-v^2q^2+i\omega/\tau=0$.  The solution of this quadratic
equation has the form
\begin{equation}
  \label{eq:crossover_second_sound_spinless}
  \omega=\pm\sqrt{(v q)^2-\frac{1}{(2\tau)^2}}-\frac{i}{2\tau},
\end{equation}
analogous to that of Eq.~(\ref{eq:crossover_second_sound}) for the
second sound in a quantum liquid of spin-$\frac12$ fermions.  Again,
at $q\ll (v\tau)^{-1}$ we recover the general expression
(\ref{eq:diffusive_solution}) for the heat diffusion mode to leading
order at $T\to0$.  For $q$ in the parametrically broad region
$(v\tau)^{-1} \ll q\ll q^*$ or, equivalently,
\begin{equation}
  \label{eq:intermediate_region}
  \tau^{-1}\ll \omega\ll \tau^{-1}\frac{v}{\delta c},
\end{equation}
Eqs.~(\ref{eq:hybrid_vs_dissipation_below_q*}) and
(\ref{eq:crossover_second_sound_spinless}) are both applicable, and
for the right-moving sound mode we have $\omega=vq-i/2\tau$.

We now summarize our results for the crossover in a spinless quantum
liquid from the regime of two hybrid sound modes at
$\omega\gg\tau^{-1}$ to the regime of ordinary sound at $\omega\to0$.
The simple description of the two sound modes with speeds
(\ref{eq:v_pm}) applies not in the whole frequency range
(\ref{eq:frequency_range}), but at
\begin{equation}
  \label{eq:frequency_range_new} 
  \tau^{-1}\frac{v}{\delta c} \ll \omega \ll \tau_{\rm ex}^{-1}.
\end{equation}
At frequencies near $\tau^{-1}(v/\delta c)$ or $q\sim q^*$ the nature
of the sound modes changes.  By combining
Eqs.~(\ref{eq:rho_eqn_linearized_with_tau}) and
(\ref{eq:hybrid_vs_dissipation}) we obtain the ratio of the variations
of density and entropy
\begin{equation}
  \label{eq:drho/dsigma}
  \frac{\delta\rho}{\delta\sigma}
   = i\sqrt{\frac{A_{12}}{A_{21}}}\,
      \frac{q^*}{q}
      \left[1\pm\sqrt{1-\left(\frac{q}{q^*}\right)^2}\,\right].
\end{equation}
In the frequency range (\ref{eq:frequency_range_new}) we recover the
ratio $\delta\rho/\delta\sigma=\pm\sqrt{A_{12}/A_{21}}$ for the two
hybrid sound modes that approaches a finite value at $T\to0$.  At $q$
below $q^*$ the oscillations of $\rho$ and $\sigma$ acquire a phase
shift $\pi/2$, while the ratio $|\delta\rho/\delta\sigma|$ depends on
$q$.  Near the lower end of the range (\ref{eq:intermediate_region})
we obtain $|\delta\rho/\delta\sigma|\sim \sqrt{A_{12}/A_{21}}\,
(T/D)^{\mp1}$, i.e., the two modes are similar to the first and second
sound in the liquid of spin-$\frac12$ fermions.  Finally, at
$\omega\ll\tau^{-1}$ the first sound mode becomes the ordinary sound,
whereas the second one is replaced by heat diffusion.

\section{Attenuation of sound due to the fast scattering processes}
\label{sec:attenuation_fast}

The two sound modes in one-dimensional quantum liquids exist only in
the frequency range (\ref{eq:frequency_range}).  The limitation on the
high frequencies was imposed in Eq.~(\ref{eq:frequency_range}) in
order to make sure that the gas of excitations is always near
equilibrium, which is achieved at the time scale $\tau_{\rm ex}$.
Because the relaxation processes are not instantaneous, they lead to
attenuation of sound.  In the two-fluid hydrodynamic description of
the system this effect is accounted for by the dissipative
corrections to the currents $j_s$, $j_p$, and $j_{u_0}$ in
Eq.~(\ref{eq:currents}).  The strength of the dissipative processes in
the quantum liquid is characterized by the bulk viscosities $\zeta_i$
and the thermal conductivity of the gas of excitations $\kappa_{\rm
  ex}$, which are all proportional to $\tau_{\rm ex}$.  We now study
the effect of these processes on the attenuation of the sound modes by
assuming that $\tau_{\rm ex}$ is finite, while $\tau\to\infty$.

We start by rederiving Eqs.~(\ref{eq:rho_eqn}) and
(\ref{eq:sigma_eqn}), while accounting for the nonvanishing $\zeta_i$
and $\kappa_{\rm ex}$ in Eq.~(\ref{eq:currents}).  A straightforward
calculation yields
\begin{subequations}
  \label{eq:dissipative_equations}
\begin{eqnarray}
  \label{eq:rho_eqn_dissipative}
  \partial_t^2\rho &=& \partial_x^2\Pi 
           -\zeta_1\rho_0\partial_x^3(u_0-u_{\rm ex})
           -\zeta_2\partial_x^3u_{\rm ex},
\\[1ex]
  \label{eq:sigma_eqn_dissipative}
  \partial_t^2\sigma &=&
     \frac{\rho_0^{}}{\rho_{\rm ex}}\sigma^2\partial_x^2 T
     -\frac{\rho_0^{2}\sigma}{\rho\rho_{\rm ex}}
           (\zeta_1-\rho\:\!\zeta_3)\partial_x^3(u_0-u_{\rm ex})
\nonumber\\
&&
     -\frac{\rho_0^{}\sigma}{\rho\rho_{\rm ex}}
           (\zeta_2-\rho\:\!\zeta_4)\partial_x^3u_{\rm ex}
     +\frac{\kappa_{\rm ex}}{\rho T}\partial_{t}\partial_{x}^2T.
\end{eqnarray}
\end{subequations}
Our goal is to account for the dissipation caused by nonvanishing
viscosity and thermal conductivity in the first order in $\zeta_i$ and
$\kappa_{\rm ex}$.  Therefore we replace the derivatives of velocities
$u_0$ and $u_{\rm ex}$ in Eqs.~(\ref{eq:rho_eqn_dissipative}) and
(\ref{eq:sigma_eqn_dissipative}) by the expressions obtained from
dissipationless hydrodynamic equations (\ref{eq:dissipationless_rho})
and (\ref{eq:dissipationless_sigma}),
\begin{equation}
  \label{eq:dxuex}
  \partial_x u_{\rm ex}=-\frac{\partial_t\rho}{\rho}
                        -\frac{\partial_t\sigma}{\sigma},
\quad
  \partial_x(u_{0}-u_{\rm ex}) = \frac{\rho}{\rho_0^{}} 
                                \frac{\partial_t\sigma}{\sigma}.
\end{equation}
This results in a system of two equations for two variables: $\rho$ and
$\sigma$,
\begin{subequations}
  \label{eq:dissipative_equations_approx}
\begin{eqnarray}
  \label{eq:rho_eqn_dissipative_approx}
  \partial_t^2\rho &=& \partial_x^2\Pi 
           +\frac{\zeta_2}{\rho}\partial_t\partial_{x}^2\rho
           +\frac{\zeta_2-\rho\zeta_1}{\sigma}\partial_t\partial_{x}^2\sigma,
\\[1ex]
  \label{eq:sigma_eqn_dissipative_approx}\hspace{-2em}
  \partial_t^2\sigma &=&
     \frac{\rho_0^{}}{\rho_{\rm ex}}\sigma^2\partial_x^2 T
     +\frac{\rho_0^{}\sigma}{\rho^2\rho_{\rm ex}}
           (\zeta_2-\rho\:\!\zeta_1)\partial_t\partial_{x}^2\rho
\nonumber\\
&&
     +\frac{\rho_0^{}}{\rho\rho_{\rm ex}}\,\widetilde\zeta\,
           \partial_t\partial_{x}^2\sigma
     +\frac{\kappa_{\rm ex}}{\rho T}\,\partial_{t}\partial_{x}^2T.
\end{eqnarray}
\end{subequations}
Here we have introduced
\begin{equation}
  \label{eq:zeta_tilde}
  \widetilde\zeta=\zeta_2-2\rho\:\!\zeta_1+\rho^2\:\!\zeta_3
\end{equation}
and applied the Onsager relation $\zeta_4=\zeta_1$.

We now substitute into Eqs~(\ref{eq:dissipative_equations_approx})
small variations of $\rho$ and $\sigma$ in the form $\delta\rho\,
e^{i(qx-\omega t)}$ and $\delta\sigma\, e^{i(qx-\omega t)}$ and obtain
the following system of two linear equations
\begin{equation}
  \label{eq:system_linear}
  \Big(\omega^2-q^2\hat A+i\omega q^2\hat\alpha\Big)
  \left(
    \begin{array}{c}
      \delta\rho
    \\[1ex]
      \delta\sigma
    \end{array}
  \right)
 =  \left(
    \begin{array}{c}
      0
    \\[1ex]
      0
    \end{array}
  \right),
\end{equation}
where $\hat A$ and $\hat\alpha$ are $2\times2$ matrices, with the
matrix elements $A_{ij}$ given by Eq.~(\ref{eq:A}) and
\begin{subequations}
  \label{eq:alpha}
\begin{eqnarray}
  \label{eq:alpha11}
  \alpha_{11}^{}&=&\frac{\zeta_2}{\rho},
\\
  \label{eq:alpha12}
  \alpha_{12}^{}&=&\frac{\zeta_2-\rho\zeta_1}{\sigma},
\\
  \label{eq:alpha21}
  \alpha_{21}^{}&=&\frac{\rho_0^{}\sigma}{\rho^2\rho_{\rm ex}}
                    (\zeta_2-\rho\zeta_1)
           +\frac{\kappa_{\rm ex}}{\rho T}
          \left(\frac{\partial T}{\partial\rho}\right)_\sigma,
\\
  \label{eq:alpha22}
  \alpha_{22}^{}&=&\frac{\rho_0^{}}{\rho\rho_{\rm ex}}\widetilde\zeta
           +\frac{\kappa_{\rm ex}}{\rho T}
          \left(\frac{\partial T}{\partial\sigma}\right)_\rho.
\end{eqnarray}
\end{subequations}
The frequencies of the sound modes as functions of $q$ are obtained by
equating the determinant of the matrix in the left-hand side of
Eq.~(\ref{eq:system_linear}) with zero.  

In the leading order at $q\to0$ the term proportional to $\hat\alpha$
in Eq.~(\ref{eq:system_linear}) can be neglected, and the solutions
take the form $\omega^2=c_\pm^2q^2$, with the speeds given by
Eq.~(\ref{eq:v^2}).  Accounting for $\hat\alpha$ in the first order,
we obtain a small imaginary part of $\omega$, which for the
right-moving waves takes the form
\begin{equation}
  \label{eq:omega_correction}
  \omega=c_\pm q -i\frac{q^2}{2}\gamma_\pm,
\end{equation}
where 
\begin{eqnarray}
  \label{eq:gamma}
  \gamma_\pm&=&\frac{\alpha_{11}^{}+\alpha_{22}^{}}{2}
            \pm\frac{(A_{11}-A_{22})(\alpha_{11}^{}-\alpha_{22}^{})}
                    {2\sqrt{(A_{11}-A_{22})^2+4A_{12}A_{21}}}
\nonumber\\
          &&\pm\frac{A_{12}\alpha_{21}^{}+A_{21}\alpha_{12}^{}}
                    {\sqrt{(A_{11}-A_{22})^2+4A_{12}A_{21}}}.
\end{eqnarray}
We now analyze the sound attenuation given by
Eqs.~(\ref{eq:omega_correction}) and (\ref{eq:gamma}) for
one-dimensional quantum liquids at low temperatures.

In the case of the liquid of spin-$\frac12$ fermions at $T\to0$ the
matrix elements $A_{12}$ and $A_{21}$ scale linearly with $T$, whereas
the difference $A_{11}-A_{22}$ approaches a finite positive value.  As
a result, for the attenuation of the first sound we get
\begin{subequations}
  \label{eq:gamma_pm_spins}
\begin{eqnarray}
  \label{eq:gamma+}
  \gamma_+&=&\alpha_{11}^{}
           +\frac{A_{12}A_{21}\alpha_{22}^{}}{(A_{11}-A_{22})^2}
           +\frac{A_{12}\alpha_{21}^{}+A_{21}\alpha_{12}^{}}{A_{11}-A_{22}}.
\nonumber\\
  \label{eq:gamma+final}
         &=&\frac{\zeta_2}{\rho}
           +2\frac{\zeta_2-\rho\:\!\zeta_1}{\rho}
            \frac{\bar v^2\partial_\rho(\rho \widetilde v)}
                 {\widetilde v(v_\rho^2-\bar v^2)}
           +\frac{\widetilde\zeta}{\rho}
            \frac{\bar v^4[\partial_\rho(\rho \widetilde v)]^2}
                 {\widetilde v^2(v_\rho^2-\bar v^2)^2}
\nonumber\\
&&+T\kappa_{\rm ex}
            \frac{v_\rho^2[\partial_\rho(\rho \widetilde v)]^2}
                 {\rho\widetilde v^2(v_\rho^2-\bar v^2)^2}.
\end{eqnarray}
Here we have evaluated the coefficients next to $\zeta_1$, $\zeta_2$,
$\widetilde \zeta$, and $\kappa_{\rm ex}$ to leading order at $T\to0$.
Unlike the result (\ref{eq:gamma+}), the attenuation of the second
sound is dominated by just one of the four matrix elements of
$\hat\alpha$,
\begin{equation}
  \label{eq:gamma-}
 \gamma_-=\alpha_{22}=
           \widetilde\zeta\,
           \frac{3\hbar\widetilde v\bar v^2}{\pi T^2}
           +\kappa_{\rm ex}
           \frac{3\hbar\widetilde v}{\pi T}.
\end{equation}
\end{subequations}
Note, that the decay rate of the first sound is much smaller than that
of the second sound, $\gamma_+/\gamma_-\sim (T/D)^2$.  This is a
result of the fact that the dissipation occurs in the gas of
excitations, which is much more strongly disturbed by the second sound
than the first one.

For a spinless quantum liquid we have $(A_{11}-A_{22})^2\ll
A_{12}A_{21}$.  Then using the expressions (\ref{eq:alpha}) we
conclude that the two hybrid modes decay at the same rate, such that
in Eq.~(\ref{eq:omega_correction}) $\gamma_\pm=\alpha_{22}^{}/2$.
This result takes the form
\begin{equation}
  \label{eq:gamma_pm_spinless}
  \gamma_\pm=\widetilde\zeta\,
           \frac{3\hbar v^3}{2\pi T^2}
           +\kappa_{\rm ex}
           \frac{3\hbar v}{2\pi T}
\end{equation}
in terms of the parameters of the quantum liquid.

\section{Summary and discussion of the results}
\label{sec:discussion}

Relaxation of one-dimensional quantum liquids is characterized by two
very different time scales $\tau$ and $\tau_{\rm ex}$.  As a
consequence, in a broad range of frequencies
(\ref{eq:frequency_range}) the quantum liquid behaves as a superfluid
and supports two sound modes.  The properties of these two modes
depend on the microscopic nature of the quantum liquid.  In a liquid
of one-dimensional spin-$\frac12$ fermions, charge and spin
excitations propagate at two different velocities $v_\rho$ and
$v_\sigma$.  As a result the two sound modes also propagate at
different velocities, which in the low-temperature limit are $v_\rho$
and $\bar v$ given by Eq.~(\ref{eq:v_bar}).  Their nature is
essentially the same as that of the first and second sounds in
superfluid $^4$He, with the former being predominantly a wave of
density and the latter---a wave of entropy.  In a spinless
one-dimensional quantum liquid all low-energy elementary excitations
propagate at the same velocity $v$.  In this case both sound modes
propagate with speed $v$ in the zero temperature limit, but at finite
$T$ the speeds split according to Eq.~(\ref{eq:v_pm}).  These are
hybrid modes, which are fundamentally different from the first and
second sounds.  They are combined oscillations of density and entropy,
either in phase or with the phase shift $\pi$.

To study the propagation and attenuation of sound in one-dimensional
quantum liquids we have adapted the two-fluid hydrodynamic theory
developed for superfluid $^4$He \cite{landau_theory_1941,
  khalatnikov_introduction_2000} to one dimension.  In addition, we
accounted for the processes of slow relaxation of the system to
thermodynamic equilibrium, which are absent in true superfluids.  The
resulting theory describes the properties of the fluid at frequencies
$\omega\ll \tau_{\rm ex}^{-1}$ as long as the deviations from
equilibrium are small.  It enabled us to study attenuation of sound in
the single-fluid regime $\omega\ll\tau^{-1}$, two-fluid regime
(\ref{eq:frequency_range}), and in the crossover region between them.
The crossover is described in detail in
Secs.~\ref{sec:crossover_spins} and \ref{sec:crossover_spinless}.  We
point out that in the case of spinless quantum liquid the crossover
splits into two: one at $\omega\sim\tau^{-1}$ and the other at
$\omega\sim\tau^{-1}v/\delta c$.  The intermediate frequency regime
(\ref{eq:intermediate_region}) arises due to the presence of a large
parameter $v/\delta c \sim D/T$; it occupies a small part of the
exponentially broad region (\ref{eq:frequency_range}).  In the range
(\ref{eq:intermediate_region}) the dissipation processes significantly
affect the nature of the sound modes, transforming hybrid sounds into
modes similar to the first and second sound.

We then studied the effect on sound attenuation of the fast relaxation
processes occurring on the time scale $\tau_{\rm ex}$.  The resulting
sound attenuation is described by the last term in
Eq.~(\ref{eq:omega_correction}).  In the frequency range
(\ref{eq:frequency_range}) this contribution may compete with the
attenuation caused by the slow relaxation processes.  The total
attenuation rate is obtained by adding $-i(q^2/2)\gamma_\pm$ to
Eqs.~(\ref{eq:attenuation_high_omega_spins}) and
(\ref{eq:attenuation_high_omega_hybrid}), with $\gamma_\pm$ given by
Eqs.~(\ref{eq:gamma_pm_spins}) and (\ref{eq:gamma_pm_spinless}),
respectively.  It is worth noting that the different physical nature
of the sound waves for systems with and without spins results in
qualitatively different sound attenuation.  Because the latter is
enabled by the gas of excitations, the first sound, being
predominantly a density wave, is weakly damped compared with the
second sound in a liquid of spin-$\frac12$ fermions.  On the other
hand, the gas of excitations is affected equally by both hybrid modes
in a spinless system, resulting in equal attenuation rates.

The phenomenological parameters $\zeta_1$, $\zeta_2$, $\zeta_3$, and
$\kappa_{\rm ex}$ are expected to be proportional to the fast
relaxation time $\tau_{\rm ex}$.  However, at this time a microscopic
expression has been obtained \cite{matveev_viscous_2017,
  degottardi_viscosity_2018} only for $\zeta_2$, which coincides with
the bulk viscosity in the single-fluid regime $\omega\to0$.
In a single-channel liquid of one-dimensional spinless fermions, the
temperature dependence of the bulk viscosity is given by
$\zeta_2\propto T^4\tau_{\rm ex}$ \cite{matveev_viscous_2017}.
Assuming that all $\zeta_i$ have the same temperature dependence, we
expect the viscous contributions to $\gamma_\pm$ given by the first
term in Eq.~(\ref{eq:gamma_pm_spinless}) to scale as $T^2\tau_{\rm
  ex}$.  Although no calculations of $\kappa_{\rm ex}$ are available
at this time, we expect the contribution of the thermal conductivity
of the gas of excitations to sound attenuation to be of the same order
as that of viscous dissipation.

One-dimensional quantum liquids of interacting electrons can be
studied experimentally in long quantum wire devices, such as that of
Ref.~\cite{yacoby_nonuniversal_1996}.  Alternatively, atoms confined
in elongated traps \cite{kinoshita_observation_2004,
  moritz_confinement_2005} can also form a one-dimensional quantum
liquid.  To achieve quantum regime, the system must be cooled well
below the bandwidth $D$, which is usually of the order of the Fermi
energy for systems of fermions.  The observation of sound modes
requires temperatures that are low enough for the exponentially small
rate $\tau^{-1}$ to become much smaller than the power-law rate
$\tau_{\rm ex}^{-1}$.  One of the more favorable systems in this
respect is that of weakly interacting spin-$\frac12$ fermions, for
which $\tau_{\rm ex}^{-1}\propto T$ \cite{karzig_energy_2010}, as
opposed to $\tau_{\rm ex}^{-1}\propto T^7$ in spinless systems
\cite{arzamasovs_kinetics_2014, protopopov_relaxation_2014}.  In the
temperature regime considered in Ref.~\cite{karzig_energy_2010} the
interaction corrections to the Fermi velocity of electrons are small,
and the collective charge and spin excitations are not formed.  In
this case the Fermi velocity is the only speed of low-energy
excitations in the system, despite the presence of spins.  As a
result, we expect this system to have properties similar to those of
spinless quantum liquids, including two hybrid sound modes.

\begin{acknowledgments}

  Work at Argonne National Laboratory was supported by the
  U.S. Department of Energy, Office of Science, Materials Sciences and
  Engineering Division.  Work at the University of Washington was
  supported by the U.S.  Department of Energy Office of Science, Basic
  Energy Sciences under Award No. DE-FG02-07ER46452.

\end{acknowledgments}


\begin{thebibliography}{32}
\bibitem [{one()}]{onefootnote}%
  \BibitemOpen
  \href@noop {} {}\bibinfo {note} {Henceforth we limit our discussion to
  right-moving waves.}\BibitemShut {Stop}%
\bibitem [{two()}]{twofootnote}%
  \BibitemOpen
  \href@noop {} {}\bibinfo {note} {Strictly speaking the square-root
  singularity in Eq.~(\ref{eq:hybrid_vs_dissipation}) at $q=q^*$ is smeared
  slightly by $\delta q^*\sim (v\tau)^{-1}$. This smearing can be obtained by
  substituting $\omega=vq-i/4\tau$ instead of $\omega=vq$ into the right-hand
  side of Eq.~(\ref{eq:quadratic_eq_solution}).}\BibitemShut {Stop}%
\end{thebibliography}
\end{document}